# Self-Propelling Rotator Driven by Soluto-Capillary Marangoni Flows[1]

Mark Frenkel[2], Gene Whyman[2], Evgeny Shulzinger[2], Anton Starostin[3],

Edward Bormashenko[1,a)]

[1]Ariel University, Engineering Faculty, Chemical Engineering Department, P.O.B. 3, 40700, Ariel, Israel

[2]Ariel University, Natural Sciences Faculty, Physics Department, P.O.B. 3, 40700, Ariel, Israel

[3]Institute of Technical Chemistry, Ural Division, Russian Academy of Science, Academika Koroleva Street 3, Perm 614013, Russia

[a]Corresponding author: Edward Bormashenko, E-mail: edward@ariel.ac.il

ABSTRACT

The self-propelled, longstanding rotation of the polymer tubing containing camphor continuing for dozens of hours is reported. The rotator is driven by the solutocapillary Marangoni flows owing to the dissolution of camphor. The phenomenological model of self-propulsion is suggested and verified. Scaling laws describing the quasi-stationary self-propulsion are proposed and tested experimentally. The change in the surface tension, arising from the dissolution of camphor and driving the rotator is estimated as 0.3 mN/m.

---

[1] Was published in APPLIED PHYSICS LETTERS  Volume: 110  Issue: 13  Article Number: 131604   Published: MAR 27 2017

Autonomous displacement of nano-, micro- and macroscopic objects, containing their own means of motion, called also self-locomotion (or self-propulsion) and driven mainly by interfacial phenomena, attracted considerable attention from investigators.[1–6] Various mechanisms of self-propulsion have been introduced and investigated, which exploited phoretic effects[7], gradient surfaces, breaking the wetting symmetry of a droplet on a surface[8-11], the Leidenfrost effect[12-18], the self-generated hydrodynamic and chemical fields, originating from the geometrical confinements[19-20], and soluto- and thermo-capillary Marangoni flows.[21-27] Within self-propelling systems, micro-rotators are distinguished, which are of a special interest due to the hydrodynamic coupling effects inherent to these systems[28-32]. Our paper is devoted to theoretical and experimental investigations of the self-propelled rotator driven by the camphor engine. Dissolution of camphor in water gives rise to the Marangoni soluto-capillary flow resulting in the self-propulsion of so-called camphor boats, which were subject to the intensive research recently[33-36].

Polyvinyl chloride (PVC) tubing was used for the manufacturing of the rotator as shown in Fig. 1. The diameter of the tubing was $2R = 2.5$ mm, the thickness was 0.8 mm, the length $2L$ was varied in the range from 5.5 to 61 mm (7 samples, see Fig 4); the mass of the PVC tubing $m_{tub}$ was varied from 0.03 to 0.4 g. Camphor (96%), supplied by Sigma–Aldrich, was placed at the ends of the PVC tubing, as shown in Fig. 1. The similar construction of self-propelling system exploiting the Marangoni soluto-capillary flow has been reported recently in Ref. 37.

Bi-distilled water (the specific resistivity $\rho \cong 18 \cdot M\Omega \cdot cm$ at room temperature) was used as a supporting liquid. The mass of camphor $m_{cam}$ was $ca$ 0.005 g. It is seen, that the condition $m_{tub} \gg m_{cam}$ takes place for all lengths of rotators; thus $m \cong m_{tub}$ occurs, where $m$ is the total mass of the rotator. Supporting water was placed into the Petri dish with the diameter of 190 mm. The experiments were carried at ambient conditions ($t = 21^0 C$; relative humidity $RH = 35 - 40\%$).

The motion of the boat was recorded by CASIO Digital Camera EX-FH20. After capturing the video, the movie was split into separate frames by VirtualDub software. The video frames were treated by the specially developed homemade software, which allowed calculation of the speed of the rotator.

When PVC tubing containing a "camphor engine" was placed onto the water surface, it started to rotate, as depicted in Fig. 2 (Multimedia view). Surprisingly, rotation may continue for dozens of hours. The time dependence of the angular velocity of the rotator is supplied in Fig. 3. The timespan of quasi-stationary rotation, when the angular velocity $\omega$ is only slightly time dependent ($\omega \cong const$) is clearly recognizable.

Rotation of the tubing is obviously due to the jump in the surface tension, $\Delta \gamma$, between two sides across the tubing, which is conditioned by the special shape of the tubing ends cut parallel under the angle of 45° (see Fig. 1). Such a shape of the tubing ends provided the gradients of camphor concentration in opposite directions along two sides of the tubing.

For the deriving of simple scaling relations, assume $\Delta \gamma = const$ for the quasi-stationary rotation of the tubing, expected when the driving force arising from the change in the surface tension is balanced by the viscosity force. Thus, the moment of the driving force rotating the tubing $M^\gamma$ is calculated as (the frame of reference is shown in Fig. 1):

$$M^\gamma = 2\int_0^L \Delta \gamma x dx = \Delta \gamma L^2 , \qquad (1)$$

This moment is balanced by the moment of viscous force within the stationary timespan. The viscous stress $\sigma^{visc}$ may be roughly estimated as follows:

$$\sigma^{visc} \propto \eta \frac{v}{R} = \eta \frac{\omega x}{R}, \qquad (2)$$

where $v$ is the velocity of the point located on the rotator at the distance $x$ from its center, $\eta$ is the viscosity of water, $R=1.25$ mm is the external radius of the tubing. Hence, the moment of the viscous forces acting on the rotator is calculated with Eq. (3):

$$M^{visc} = 2\int_0^L R\sigma^{visc} x dx \propto 2\int_0^L \eta \omega x^2 dx = \frac{2}{3}\eta \omega L^3 \qquad (3)$$

Equating $M^{\gamma} = M^{visc}$ yields:

$$\omega \propto \frac{3}{2}\frac{\Delta \gamma}{\eta L} \qquad (4)$$

Actually, Eq. (4) already follows from the dimensional analysis of the problem and predicts the scaling law $\omega \sim 1/L$ for the angular velocity of the rotator (the multiplier 3/2 should not be taken too seriously, due to the crudeness of the estimations, resulting in Eq. 4). This scaling law was checked experimentally with the tubing of different lengths. The experimental data illustrated with Fig. 4 justify the scaling law predicted by Eq. (4); thus, validating the entire model proposed for the quasi-stationary rotation of the self-driven propeller. The expression (4) also enables the rough estimation of the jump in surface tension $\Delta \gamma \propto \omega \eta L$, driving the rotator. Assuming $\eta \approx 10^{-3}$ Pa×s for water, we obtain $\Delta \gamma \cong 0.3$ mN/m.

Now we address the time scale of stabilization of the rotation, which is on the order of magnitude of a couple of seconds (see Fig. 3). It is reasonable to relate the origin of this time scale to the viscous dissipation, the characteristic time of which may be estimated as: $\tau_v \cong m/(2\eta L) = \rho_{l,\text{boat}}/\eta$ where $\rho_{l,\text{boat}} = 6.6 \times 10^{-3}$ kg/m is the linear density of the PVC tubing; thus, we obtain $\tau_v \cong 6.6 s$ which is in a reasonable agreement with experimental data (it is noteworthy that the value of $\tau_v$ is independent on the tubing length).

A surprising feature of the rotator behavior is the timespan of its self-stabilized self-propulsion, as long as a dozen of hours. Levelling of the surface tension due to dissolution of camphor does not occur for a relatively long time. It should be emphasized, that the proposed model of the quasi-stationary flow neglects the effects arising from the evaporation of camphor, diffusion of camphor into water bulk, and mixing of water due to the rotation of the tubing; hence further development of the model is planned. We observed experimentally that closing the Petri dish with the cover decreased the time of rotation by 2-3 times. This means that intensive evaporation of camphor from the water surface promotes the long-lasting self-propulsion.

The mechanism of rotation, arising from Marangoni solutocapillary flows, is similar to that governing the motion of "camphor boats".[33-39] The surface tension jump, owing to the dissolution of camphor ($\Delta\gamma \cong 0.3$ mN/m) rotating the tube is in a satisfactory agreement with the data extracted from the analysis of the motion of camphor boats.[40] The reported self-propelled rotator demonstrates a potential as a micro-mixer[41] and micro-electrical-generator.[42]

**REFERENCES**
[1] N. L. Abbott, O. D. Velev, Colloid & Interface Sci. **21**, 1–3 (2016).
[2] Al. Shapere, Fr. Wilczek, Phys. Rev. Lett. **58**, 2051-2054 (1987).
[3] J. Bico, D. Quere, J. Fluid Mech. **467**, 101- 127(2002).
[4] Ghosh, P. Fischer, Nano Lett. **9 (6)**, 2243–2245 (2009).

**Acknowledgements**

The authors are indebted to Mrs. Yelena Bormashenko for her kind help in preparing this manuscript. The research was partially supported by the Government of Perm Krai, Russia research project № C-26/004.06. Acknowledgement is made to the donors of the Israel Ministry of Absorption for the partial support of the scientific activity of Dr. Mark Frenkel.

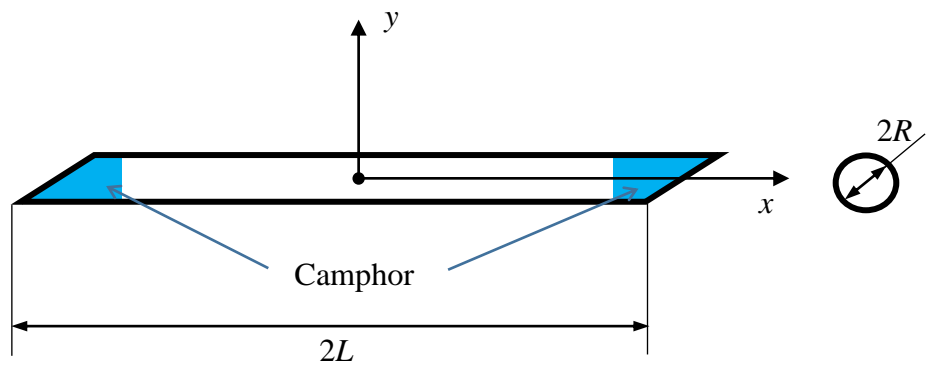

FIGURE. 1. The scheme of the rotator. Camphor is introduced at the ends of PVC tubing, which are specially cut, enabling the asymmetry of the surface tension, driving the rotator.

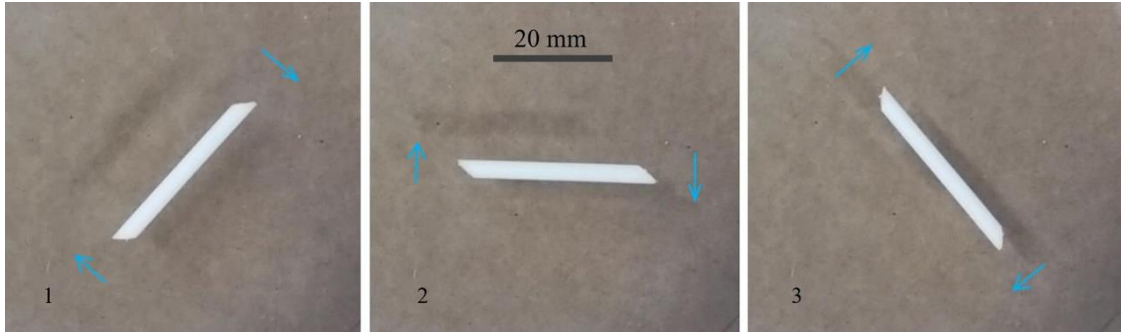

FIGURE 2. The sequence of images demonstrating the rotation of the self-propelled PVC tubing, containing camphor. The time separation between frames is 0.33 s.

(Multimedia view)

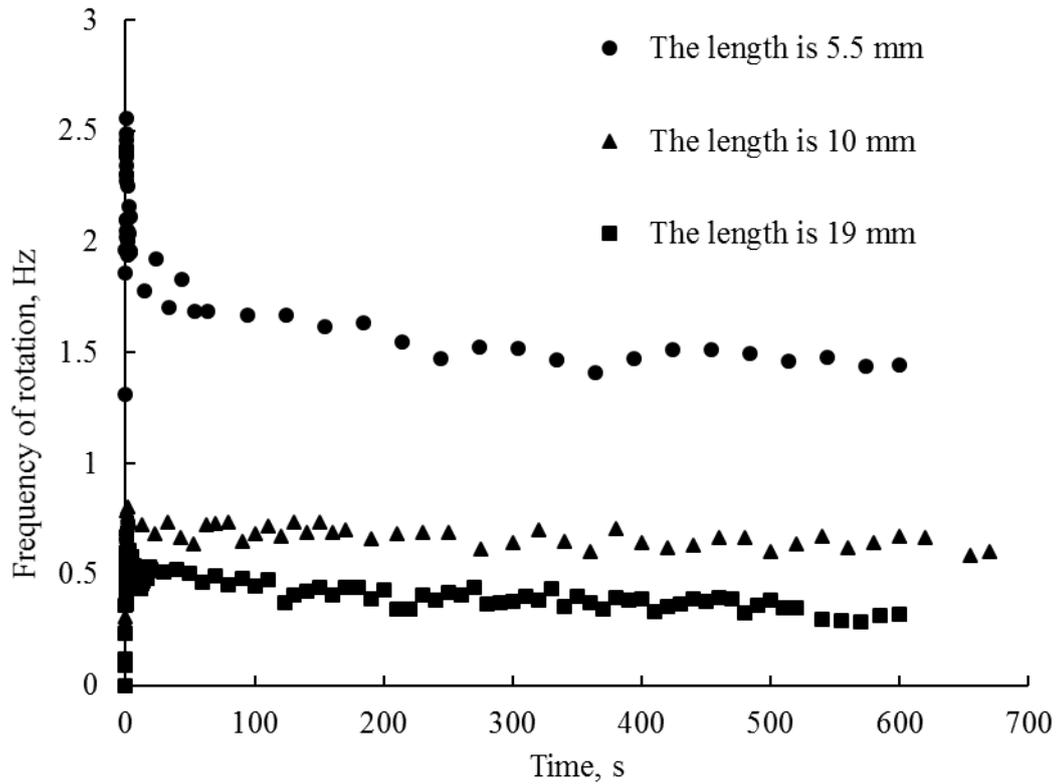

(a)

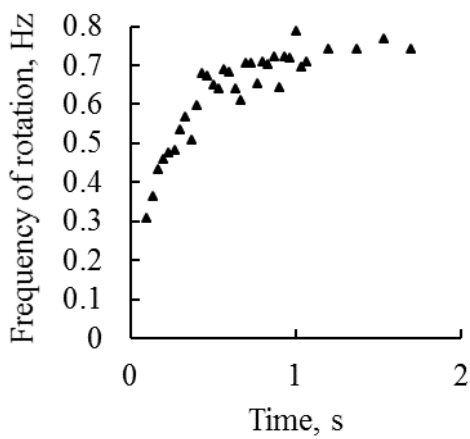

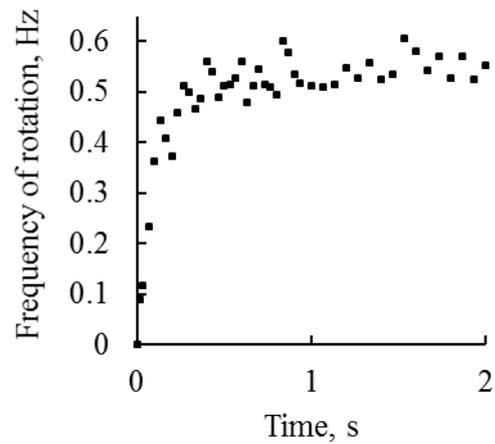

(b) (c)

FIGURE. 3. The time dependence of the frequency of rotation of the PVC tubing established for different lengths 2$L$ (a). Stabilization of the frequency is clearly seen. Time dependence at initial stages of rotation is depicted in inserts (b) for 2$L$=10 mm and (c) for 2$L$=19 mm.

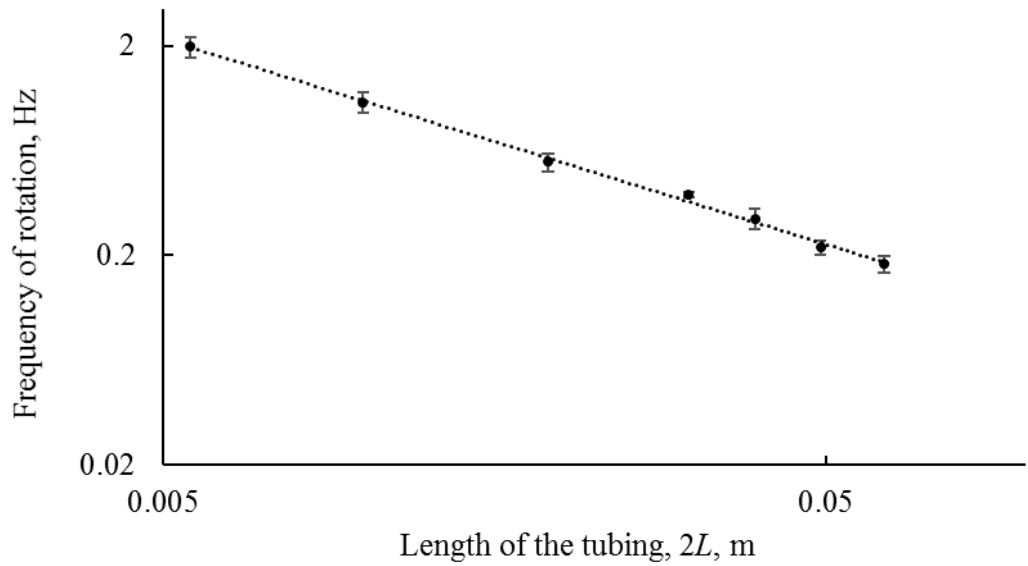

FIGURE 4. The double logarithmic plot of the frequency of rotation $f$ vs the tubing length $2L$. The fitting line is: $f = \omega/2\pi = 0.0238L^{-0.981}$ (the correlation coefficient is 0.998).